\documentclass[sigconf]{acmart}

\usepackage[colorinlistoftodos,prependcaption,textsize=tiny]{todonotes}
\usepackage{dirtytalk} 
\usepackage{listings}
\usepackage{color}

\definecolor{dkgreen}{rgb}{0,0.6,0}
\definecolor{gray}{rgb}{0.5,0.5,0.5}
\definecolor{mauve}{rgb}{0.58,0,0.82}

\AtBeginDocument{%
  \providecommand\BibTeX{{%
    \normalfont B\kern-0.5em{\scshape i\kern-0.25em b}\kern-0.8em\TeX}}}

\copyrightyear{2024}
\acmYear{2024}
\setcopyright{rightsretained}
\acmConference[ICSE-SEET '24]{46th International Conference on Software Engineering: : Software Engineering Education and Training}{April 14--20, 2024}{Lisbon, Portugal}
\acmBooktitle{46th International Conference on Software Engineering: : Software Engineering Education and Training (ICSE-SEET '24), April 14--20, 2024, Lisbon, Portugal}
\acmDOI{10.1145/3639474.3640052}
\acmISBN{979-8-4007-0498-7/24/04}

\begin{document}

\title[LLMs Still Can't Avoid Instanceof]{LLMs Still Can't Avoid Instanceof: An Investigation Into GPT-3.5, GPT-4 and Bard's Capacity to Handle Object-Oriented Programming Assignments}

\author{Bruno Pereira Cipriano}
\email{bcipriano@ulusofona.pt}
\orcid{0000-0002-2017-7511}
\affiliation{
  \institution{Lusófona University, COPELABS}
  \streetaddress{Campo Grande, 376}
  \city{Lisbon}
  \country{Portugal}
  \postcode{1700-097}
}
\author{Pedro Alves}
\orcid{0000-0003-4054-0792}
\email{pedro.alves@ulusofona.pt}
\affiliation{
  \institution{Lusófona University, COPELABS}
  \streetaddress{Campo Grande, 376}
  \city{Lisbon}
  \country{Portugal}
  \postcode{1700-097}
}

\renewcommand{\shortauthors}{Cipriano, B. Pereira and Alves, P.}

\begin{abstract}
Large Language Models (LLMs) have emerged as promising tools to assist students while solving programming assignments. However, object-oriented programming (OOP), with its inherent complexity involving the identification of entities, relationships, and responsibilities, is not yet mastered by these tools. Contrary to introductory programming exercises, there exists a research gap with regard to the behavior of LLMs in OOP contexts. In this study, we experimented with three prominent LLMs - GPT-3.5, GPT-4, and Bard - to solve real-world OOP exercises used in educational settings, subsequently validating their solutions using an Automatic Assessment Tool (AAT). The findings revealed that while the models frequently achieved mostly working solutions to the exercises, they often overlooked the best practices of OOP. GPT-4 stood out as the most proficient, followed by GPT-3.5, with Bard trailing last. We advocate for a renewed emphasis on code quality when employing these models and explore the potential of pairing LLMs with AATs in pedagogical settings. In conclusion, while GPT-4 showcases promise, the deployment of these models in OOP education still mandates supervision. 
\end{abstract}




\maketitle

\section{Introduction}

In the last few years, researchers introduced us to Large Language Models (LLMs), which are tools that can predict the next word in a sequence, after being trained on large amounts of data, and taking into consideration a large number of parameters. Popular implementations of LLMs are ChatGPT, a chatbot developed by OpenAI on top of their Generative Pre-trained Model (GPT), and GitHub Copilot\footnote{https://github.com/features/copilot}, a developer support tool that can generate code from annotations and is also based on GPT.



As CS educators, we are interested in determining how these tools can be used to support students in their programming learning process, both inside and outside of the classroom. We are specifically interested in the possibilities of using these tools for teaching and learning object-oriented design and programming.

Previous research has shown that LLMs have the capacity to generate computer program code from natural language descriptions \cite{xu2022systematic}. Furthermore, GPT-based models, such as Codex, have been shown to be able to solve most types of exercises that are used in introductory programming classes \citep{finnie2022robots}. Additionally, researchers have reported that GPT-3.5 can handle object-oriented programming assignments, managing to reach decent to good scores, with the caveat that the generated code is not up-to-par with the industry's best practices of that paradigm \citep{10.1145/3587102.3588814}.

However, both the availability and the capabilities of these tools are increasing rapidly. Updated models, such as GPT-4\footnote{Launched on March 14, 2023}, promise even better performance than their older versions \citep{openai_gpt-4_2023, bubeck2023sparks}, while new tools, such as Bard\footnote{Launched in the US \& UK on Mar 21, 2023; in EU \& Brazil on Jul 13, 2023.}, a chatbot developed by Google on top of their Language Model for Dialogue Applications (LaMDA) \citep{google-bard-message-from-ceo, thoppilan2022lamda}, are becoming available to the general public. These new models justify performing further research on these topics, to confirm and/or update the findings that resulted from the previous works.



Thus, we decided to explore these newer tools to learn their capabilities and limitations, as well as identify opportunities to integrate them into our courses, similarly to what is being done by other instructors \citep{lau2023ban, denny2023promptly, daun2023chatgpt, leinonen2023comparing, liffiton2023codehelp}.




As such, we performed an evaluation of some of the most recently available LLMs, GPT-4, and Bard, following previous research focused on object-oriented design, programming and best-practices using GPT-3.5 \citep{10.1145/3587102.3588814}. This evaluation was based on assignments that are automatically graded by an Automatic Assessment Tool (AAT).

This paper makes the following contributions:
\begin{itemize}
\item Presents and compares the performance (as measured by an AAT) of GPT-3.5, GPT-4, and Bard using 6 real-world assignments that have been used to grade students in an Object-Oriented Programming university course;
\item Identifies and categorizes the errors found in the code generated by the tested LLMs;
\item Provides full logs of the interactions used to get each model to solve one of the assignments, in an annotated prompt/reply format;
\item Presents a set of recommendations for CS Educators wishing to adapt their classes to the availability of these technologies.
\end{itemize}

\section{Related work}


In the last couple of years, a relevant body of research related with measuring LLMs' ability to handle introductory programming exercises has been published \citep{chen2021evaluating, finnie2022robots, finnie2023my,
openai_gpt-4_2023, reeves2023evaluating, wermelinger2023using, destefanis2023preliminary}. Those introductory exercises are usually solved with a single function that receives a set of parameters and transforms them into an expected output. For example, in \cite{chen2021evaluating}, researchers managed to solve 70\% of 164 Python programming exercises using code generated by Codex, a GPT-based language model trained on code publicly available in GitHub. In \citep{finnie2022robots}, Codex was evaluated using 23 introductory programming assignments, and managed to make the 17th position when ranked alongside 71 students. The authors of \citep{reeves2023evaluating} focused on Codex's capacity to handle Python-based Parsons problems \cite{parsons2006parson} and revealed that the model solved 50\% of the cases when indentation errors were considered, and 80\% when those errors were ignored. Finally, a study \citep{destefanis2023preliminary} based on single-function Java exercises from CodingBat\footnote{https://codingbat.com}, reported that GPT-3.5 solved 90.6\% of the functions, while Bard solved 53.1\%.

Some papers have also delved into GPT's capacity to handle OOP assignments, which typically require the implementation of multiple classes that interact with each other and have mutable states. In one of these studies \citep{savelka2023can}, researchers assessed the performance of two GPT-3.5 models (``text-davinci-002'' and ``text-davinci-003'') on various introductory and intermediate Python programming exercises. Some of the intermediate exercises involved OOP, and the models' solutions were evaluated using an AAT. The first model achieved a success rate of 52.9\% in the tests, while the second model scored 70.6\%. However, due to the unavailability of the exercise statements, we could not ascertain whether the type of OOP exercises was equivalent to those presented in this research paper. The same authors have updated their research by performing the evaluations using GPT-4, which reached an improved score of 82.4\% \citep{savelka2023thrilled}. In another study \citep{ouh2023chatgpt}, researchers attempted to utilize ChatGPT (versions 3.5 and 4) for solving OOP exercises. Some of these exercises involved introductory-level object usage without inheritance or polymorphism, and the chatbot managed to provide partially correct solutions. Other exercises introduced inheritance, but a UML class diagram with the solution was provided to the students, who only had to implement the corresponding code. As it was not feasible to supply the UML diagram to the model, the generated solution remained incomplete. Despite encompassing OOP exercises, this study's tasks are either very elementary or highly guided. Another study \citep{cai2023low}, proposed the utilization of LLMs as low-code tools. While directly using ChatGPT led to solutions with weak object-oriented design, they were able to achieve good results by pre-guiding the model on OOP best practices. Finally, the authors of \citep{10.1145/3587102.3588814}, evaluated GPT-3.5 (``text-davinci-003'') using both functional as well as code quality criteria and found that, although GPT-3.5 was able to pass an average of 77.63\% of unit tests, it only passed an average of 50\% of the code quality evaluations.


As far as we know, no research focused on Bard's (or LaMDA's) ability to handle object-oriented assignments has been published.

\section{The assignments}


This research was based on assignments used for student evaluation in a Computer Engineering degree. The course occurs on the 2nd curricular year of the degree and focus mostly on teaching Object-Oriented Design and Programming using the Java programming language. In this course, students are expected to learn how to implement object-oriented software solutions following the paradigm's best practices \citep{wegner1990concepts}, with a strong focus on issues like code readability, modularity and extensibility. For example, students are expected to understand the drawbacks of using type testing, such as those permitted by `instanceof', to decide program behaviors, since that technique tends to make programs harder to modify.



The course's assignments (e.g., tests and projects, among others) are evaluated using an open-source AAT, called Drop Project \citep{cipriano2022drop}. It evaluates if the student's code is respecting the assignment's requirements using teacher-defined unit tests. It is also capable of verifying if the student's code follows the expected code quality rules and guidelines.  In this course, some assignments are configured to warn the teacher about the use of certain keywords that allow the program to work while disregarding some of the aforementioned best practices. An example of this is the usage of the `instanceof' keyword or the `getClass()' function. Due to the limitations of the plugin used by the AAT for the code quality validations (Checkstyle \cite{checkstyle-2023}), some quality validations are implemented using unit tests. A case in point is verifying whether a certain class has been declared as abstract.



To evaluate the LLMs, a set of assignments that have been used in this course as mini-tests focused on inheritance and polymorphism was used. Each assignment has a different business domain, as well as different requirements. All the assignments have the following goals: identify and implement entity relationships (both composition and inheritance), implement some getters and setters, implement a non-trivial `toString()' function, and, implement some functions that have to create, query and/or manipulate objects of several classes. Note that the relationships aren't directly provided to students, who must infer which classes are the super and sub-classes. This approach sets our study apart from others where assignments give more explicit directions \citep{ouh2023chatgpt}. 



\begin{lstlisting}[caption={Partial instructions for the ITCompany assignment}, label={lst:itcompany_instructions}, basicstyle=\scriptsize\ttfamily, 
    breaklines=true,
    breakindent=0pt,
    frame=single,              
    captionpos=b, float]
We want to implement a software to help an IT company that provides ITConsultant services with the management of their employees. In this challenge, there exist two types of employees: those who pertain to the human resources management area (HRWorker), and the IT (information technology) experts, who pertain to ITConsultant.

The classes Employee, HRWorker and ITConsultant must be created (...) however it should not be possible to instantiate an object using the super class's constructor. 

A company is characterized by its name. A company can have multiple employees, but each employee only the belongs to a single company.

A employee is identified by its id (int), name (String) and monthly salary (int).
(...)
Add, to the appropriate classes, the following methods:
- A public int getId() method, which returns the employee's ID.
- A public int getSalary() method, which returns the employee's salary.
- A public int getValue() method, which returns the IT consultant's hourly rate.
(...)

(Within the Company class) add a public String toString() method, which must be implemented to return the following:
- "The company <name> does not have employees." - if the company does not have any employee;
- "The company <name> has <number_employees> employees:" - if the company has at least one employee. Besides that, the information for each employee should be display, considering the respective type.
\end{lstlisting}

Moreover, each assignment has behaviors based on the object type to assess whether students can devise solutions without resorting to explicit type checks like those allowed by the `instanceof' keyword and `getClass()' function. Finally, in some cases, we ask for the student ID to be used as the value of an object's attribute, to guarantee unique solutions among students. 






Listing~\ref{lst:itcompany_instructions} presents a partial example of an assignment's text.


Six such assignments were used in this study. The selected assignments were originally used to grade students in different school years (2018/19, 2021/22, and 2022/23).


\section{Methodology}

To interface with GPT-3.5, openAI's Playground\footnote{available at https://platform.openai.com/playground} (free account) was used. The selected model was ``text-davinci-003''. The ``temperature'' parameter, which controls randomness, was left at the default value of 0.7, and the ``maximum length'' parameter was left  at the default of 256. GPT-4 was studied using OpenAI's ChatGPT Plus\footnote{available at https://chat.openai.com/} (paid account). Finally, to study Bard, we used the respective chat interface \footnote{available at https://bard.google.com/}. Note that neither ChatGPT Plus nor Bard allows changing the models' parameters. The experiments were done in different time periods, since we did them when the models became available to us: the original experiment with GPT-3.5 was done in December 2022, the experiment with GPT-4 was done in May 2023, and finally, the experiment with Bard was done in July 2023. To evaluate the models' output, version 0.9.7 of the Drop Project AAT was used.


For each model and assignment, the following algorithm was performed:
\begin{enumerate}
    \item paste the assignment’s text into the LLM’s text input area
    \item submit it
    \item examine the output to determine whether all the mandatory classes and functions are present
    \begin{enumerate}
        \item if they are, move to step 4
        \item otherwise, prompt for the missing code and repeat step 3
    \end{enumerate}
    \item inspect the code for syntactic, logic, and output format errors
    \item manually fix any syntactic / compilation errors
    \item submit the LLM's generated code to the AAT
    \item analyze the AAT’s output
\end{enumerate}


\section{It works but it's not OO - how LLM(s) tried to solve the IT Company assignment}
\label{section:our_experience}
\begin{figure}[ht]
\centering
\includegraphics[width=1\linewidth]{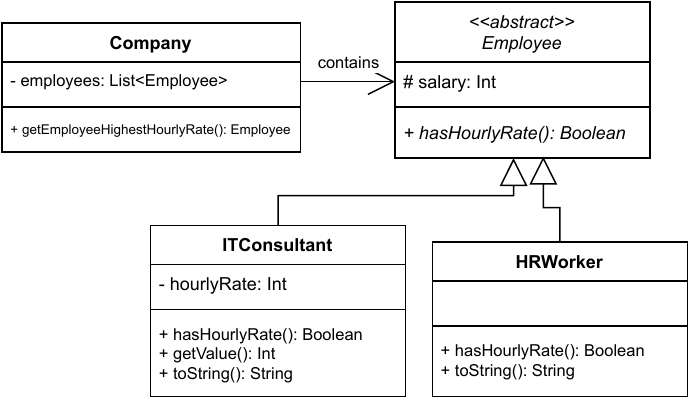}
\caption{Simplified UML class diagram for the ``IT Company'' assignment, excluding non-essential attributes, methods, and constructors. Students only receive textual instructions, not this diagram.}
\label{fig:itcompany-uml}
\end{figure}


This section presents a high level overview of our attempts to use each model to solve one of the 6 assignments: the ``IT Company'' assignment. In this assignment, students are expected to design the object model of an IT Company with the following concepts: `Company', `Employee', `ITConsultant', and `HRWorker'. The concepts and their attributes are explained to the students by text. However, the text does not explicitly indicate the relations that exist between the classes: it is up to the student to infer that there exists a composition relationship --- between the `Company' and the `Employee' classes ---, and an inheritance relationship where the `Employee' class is the super-class and the `ITConsultant' and `HRWorker' classes are the sub-classes. Students are also expected to understand that the `Employee' class should be abstract, since the text mentions that it must not be possible to instantiate that class. Finally, students must implement several mandatory functions, identifying the class of the hierarchy where each function belongs. For example, there is an attribute that only makes sense for the `ITConsultant' class: the hourly rate. As such, only that class should have that attribute and the respective getter, which must be called `getValue()'. Also mandatory is the creation of a function called `getEmployeeHighestHourlyRate()` in the `Company' class. This function must return the `Employee' with the highest hourly rate. However, since the hourly rate concept only applies to the `ITConsultant' class, the student must find the proper Object-Oriented design to implement this challenge. As such, if a student uses `instanceof' or `getClass()', a penalty will be applied to the respective grade. Figure~\ref{fig:itcompany-uml} presents an overview of this assignment’s required classes, as well as the most relevant design expectations. Listing~\ref{lst:itcompany_instructions} presents part of this assignment's instructions. A full version of the assignment, including both instructions as well as unit tests is available online.\footnote{https://github.com/drop-project-edu/itCompanyAssignment} Refer to \citep{anonymous_2023_8246165} for full logs of these interactions.


%
%



\subsection{Main analysis}



\textbf{GPT-3.5} required \textbf{18} interactions to generate a solution that included all mandatory elements. Issues found in that solution:
\begin{itemize}
    \item Function `getEmployeeHighestHourlyRate()' uses `instanceof'.
    \item Fails 1 test because the `Employee' class is not abstract.
    \item Fails 2 tests because the package name was incorrectly included in `ITConsultant.toString()' function's return value.
    \item Fails the test for the `Main.myCompany()' function due to using an uninformed number in the solution.\footnote{\label{expected-failure}This problem was expected, since we did not input the student ID to the models.}
    \item Failed to import the `ArrayList' class.
\end{itemize}



\textbf{Bard} required \textbf{11} interactions to generate a solution that included all mandatory elements. Issues found in that solution:
\begin{itemize}
    \item Function `getEmployeeHighestHourlyRate()' uses `instanceof'.
    \item Fails 1 test because the `Employee' class is not abstract.
    \item Fails 2 tests due to incorrect usage of the downcast operator in the `getEmployeeHighestHourlyRate()'.
    \item Fails 2 tests due to not respecting the expected format of the `HRWorker.toString()' and `Company.toString()' functions.
    \item Fails the same test related to the `Main.myCompany()' function that the other models failed, with a difference in the implementation: instead of guessing a number, Bard's code attempts to obtain the student ID by using `System.getEnv()', resulting in a `NullPointerException'.\footnote{See footnote \ref{expected-failure}}
    \item Failed to import the `ArrayList' class.
    \item Declared 3 attributes of the `Employee' class as private, and tried to access them directly in the `ITConsultant' sub-class.
\end{itemize}




\textbf{GPT-4} required \textbf{4} interactions to generate a solution that included all mandatory elements. Issues found in that solution:
\begin{itemize}
    \item Function `getEmployeeHighestHourlyRate()' uses `instanceof'.
    \item Fails 1 unit test because it did not fully respect the format of the `Company.toString()' function (a new line was missing).
    \item Fails the test related to the `Main.myCompany()' function due to using an uninformed number in the code.\footnote{See footnote \ref{expected-failure}}
\end{itemize}



\textbf{In summary...} These experiments show that these 3 LLMs were able to partially solve the `IT Company' assignment, although with some errors, both in terms of basic syntax (e.g. missing library importations), as well as more complex object-oriented programming errors (e.g. not declaring a class as abstract, or using explicit type testing to decide program behaviors). This is similar to what happened in the other 5 assignments, as described in Section~\ref{sec:overall_findings}.

\subsection{Attempts to improve the LLM(s)' solutions}

After obtaining each model's solutions for the ``IT Company'' assignment and verifying that all of them were breaking one of our code quality rules --- the restriction not to use `instanceof' ---, we attempted to guide the models toward better solutions, similar to how we expect a student would do after seeing the AAT's feedback, and to get a better grade. Note that these extra prompts are not part of our main methodology and are not represented in Table~\ref{tab:evaluation}.

These attempts to improve the models' solutions were done using an ad-hoc methodology where we identified problems with each model's solution, and attempted to give it a direct indication of the problem, to see what it would do. The first extra prompt was the same for all models: `Change the code to not use instanceof'. Afterward, we re-prompted based on what each model gave us. The prompts for GPT-3.5 and Bard ended up being similar because the respective intermediate solutions had similar problems from the object-oriented perspective: they both were assuming that all sub-classes of `Employee' have the `getValue()' function, which is not coherent with the rest of the respective solutions.

Besides the initial extra prompt, \textbf{GPT-3.5} required two more prompts to reach an acceptable solution: as a reply to the third prompt, it suggested adding the `getValue()' function to both the super-class and the sub-class with an implementation that would return 0. In this case, 0 is being used as a flag to indicate that the object should be ignored, which is coherent with the model's implementation of `getEmployeeHighestHourlyRate()'. Although that solution is more OO, it is somewhat risky: such a design would be problematic if a function `getEmployeeLowestHourlyRate()' is needed at some point in the future.


\textbf{Bard} had a behavior similar to that of GPT-3.5, although it required a fourth prompt to reach a solution acceptable from the OOP perspective. Bard's fourth solution was equivalent to GPT-3.5's third solution, thus having the equivalent problems. Note that, although the OO solution is acceptable, Bard implemented the `getValue()' function incorrectly: it is returning the value of the salary instead of the value of the hourly rate. Since we were focused on getting Bard to solve the OO issue, we ignored this implementation detail.

Finally, \textbf{GPT-4} only needed the extra initial prompt to reach an acceptable OO solution. The model suggested adding an abstract version of `getHourlyRate()' to the `Employee' class, as well as adding a concrete version of that function (returning -1, which is much safer than returning 0) to the `HRWorker' class. It also gave us the code for those changes, as well as a version of the `getEmployeeHighestHourlyRate()' function that only considers objects for which `getValue()' returns more than 0.



\textbf{In summary...}
Our experiments show that it was possible and easy to guide GPT-4 towards a fully working solution that respects the object-oriented best practice of avoiding explicit type testing. Also, while the other models also managed to reach acceptable, although risky, OO solutions, they required more effort (GPT-3.5: 3 prompts; Bard: 4 prompts) and generated some minor errors (e.g. Bard's incorrect return of the salary attribute in the `getValue()' function). This extra effort is coherent with the findings from other researchers \citep{wermelinger2023using}.


\section{Overall Findings} \label{sec:overall_findings}


This section presents the 3 models' performances across the 6 assignments. As Table~\ref{tab:evaluation} shows, all models needed multiple prompts and no model passed all the tests. 

GPT-4 generally had the best performance, requiring fewer interactions than the other models, generating fewer compilation errors and, in general, having better results in the unit tests. Bard always had equal or worse unit test results than GPT-3.5, except in the ``Home Banking'' assignment, where it surpassed both GPT models.





In terms of the Code Quality validation, we observed that GPT-4 yields the best results: while GPT-3.5 and Bard respected the code quality rules in 3 out of the 6 assignments, GPT-4 did it in 4 out of 6 assignments. Listing~\ref{lst:sample-bard-code} shows an example of the type of code that these models tend to generate: it (mostly) works, but tends to decide behaviors using explicit type testing, via `instanceof' and/or `getClass()' . In summary, similarly to what previous research found for GPT-3.5 \citep{10.1145/3587102.3588814}, the two newer models still can't avoid using `instanceof'.


It should be noted that, in the ``Condominium Mgt.'' assignment, GPT-4 managed to pass the code quality, as well as 13/14 tests. Considering that the single test failure was expected\footnote{See footnote \ref{expected-failure}}, we can consider that GPT successfully solved one of the OOP assignments.


\begin{table*}[t]
\caption{Evaluation of the 3 models' solutions for each assignment. G-3.5 is GPT-3.5. The values are related to the first solution that contained all
mandatory classes and functions. For example, GPT-3.5 needed 5 prompts to generate all classes and functions for the first assignment, that solution had 4 compilation errors, the code quality failed and it passed 8/13 unit tests.}
\label{tab:evaluation}
\begin{tabular}{l|lll|lll|lll|lll|}
\cline{2-13}
                                                        & \multicolumn{3}{c|}{\textbf{Nr. of prompts}}                                                                               & \multicolumn{3}{c|}{\textbf{Nr. of Compilation errors}}                                                                  & \multicolumn{3}{c|}{\textbf{Code quality Ok?}}                                                                              & \multicolumn{3}{c|}{\textbf{Tests passed}}                                                                                                                                                                                                                                                     \\ \hline
\multicolumn{1}{|l|}{\textbf{Assignment}}               & \multicolumn{1}{l|}{\textbf{G-3.5}}            & \multicolumn{1}{l|}{\textbf{GPT-4}}           & \textbf{Bard}             & \multicolumn{1}{l|}{\textbf{G-3.5}}           & \multicolumn{1}{l|}{\textbf{GPT-4}}           & \textbf{Bard}            & \multicolumn{1}{l|}{\textbf{G-3.5}}            & \multicolumn{1}{l|}{\textbf{GPT-4}}            & \textbf{Bard}             & \multicolumn{1}{l|}{\textbf{G-3.5}}                                                                  & \multicolumn{1}{l|}{\textbf{GPT-4}}                                                                   & \textbf{Bard}                                                                   \\ \hline
\multicolumn{1}{|l|}{Realtor Agency}                    & \multicolumn{1}{l|}{5}                         & \multicolumn{1}{l|}{2}                        & 2                         & \multicolumn{1}{l|}{4}                        & \multicolumn{1}{l|}{0}                        & 8                        & \multicolumn{1}{l|}{No}                        & \multicolumn{1}{l|}{Yes}                       & Yes                       & \multicolumn{1}{l|}{\begin{tabular}[c]{@{}l@{}}8/13\\ (61.54\%)\end{tabular}}                        & \multicolumn{1}{l|}{\begin{tabular}[c]{@{}l@{}}12/13\\ (92.31\%)\end{tabular}}                        & \begin{tabular}[c]{@{}l@{}}7/13\\ (53.85\%)\end{tabular}                        \\ \hline
\multicolumn{1}{|l|}{IT Company} & \multicolumn{1}{l|}{18} & \multicolumn{1}{l|}{4} & {11} & \multicolumn{1}{l|}{1} & \multicolumn{1}{l|}{0} & {4} & \multicolumn{1}{l|}{No} & \multicolumn{1}{l|}{No} & {No} & \multicolumn{1}{l|}{\begin{tabular}[c]{@{}l@{}}9/13\\ (69.23\%)\end{tabular}} & \multicolumn{1}{l|}{\begin{tabular}[c]{@{}l@{}}11/13\\ (84.62\%)\end{tabular}} & {\begin{tabular}[c]{@{}l@{}}7/13\\ (53.85\%)\end{tabular}} \\ \hline
\multicolumn{1}{|l|}{Home Banking}                      & \multicolumn{1}{l|}{7}                         & \multicolumn{1}{l|}{2}                        & 10                        & \multicolumn{1}{l|}{1}                        & \multicolumn{1}{l|}{0}                        & 14                       & \multicolumn{1}{l|}{Yes}                       & \multicolumn{1}{l|}{Yes}                       & Yes                       & \multicolumn{1}{l|}{\begin{tabular}[c]{@{}l@{}}7/13\\ (53.85\%)\end{tabular}}                        & \multicolumn{1}{l|}{\begin{tabular}[c]{@{}l@{}}7/13\\ (53.85\%)\end{tabular}}                         & \begin{tabular}[c]{@{}l@{}}9/13\\ (69.23\%)\end{tabular}                        \\ \hline
\multicolumn{1}{|l|}{Condominium Mgt.}                  & \multicolumn{1}{l|}{5}                         & \multicolumn{1}{l|}{2}                        & 5                         & \multicolumn{1}{l|}{12}                       & \multicolumn{1}{l|}{0}                        & 3                        & \multicolumn{1}{l|}{Yes}                       & \multicolumn{1}{l|}{Yes}                       & Yes                       & \multicolumn{1}{l|}{\begin{tabular}[c]{@{}l@{}}11/14\\ (78.57\%)\end{tabular}}                   & \multicolumn{1}{l|}{\begin{tabular}[c]{@{}l@{}}13/14\\ (92.86\%)\end{tabular}}                        & \begin{tabular}[c]{@{}l@{}}10/14\\ (71.43\%)\end{tabular}                       \\ \hline
\multicolumn{1}{|l|}{Home Cinema A/V}                   & \multicolumn{1}{l|}{5}                         & \multicolumn{1}{l|}{2}                        & 6                         & \multicolumn{1}{l|}{3}                        & \multicolumn{1}{l|}{0}                        & 2                        & \multicolumn{1}{l|}{Yes}                       & \multicolumn{1}{l|}{Yes}                       & No                        & \multicolumn{1}{l|}{\begin{tabular}[c]{@{}l@{}}8/10\\ (80\%)\end{tabular}}                           & \multicolumn{1}{l|}{\begin{tabular}[c]{@{}l@{}}9/10\\ (90\%)\end{tabular}}                            & \begin{tabular}[c]{@{}l@{}}8/10\\ (80\%)\end{tabular}                           \\ \hline
\multicolumn{1}{|l|}{Railway Co. Vehicles}              & \multicolumn{1}{l|}{11}                        & \multicolumn{1}{l|}{6}                        & 8                         & \multicolumn{1}{l|}{4}                   & \multicolumn{1}{l|}{3}                        & 10                       & \multicolumn{1}{l|}{No}                        & \multicolumn{1}{l|}{No}                        & No                        & \multicolumn{1}{l|}{\begin{tabular}[c]{@{}l@{}}11/16\\ (68.75\%)\end{tabular}}                       & \multicolumn{1}{l|}{\begin{tabular}[c]{@{}l@{}}12/16\\ (75\%)\end{tabular}}                           & \begin{tabular}[c]{@{}l@{}}10/16\\ (62.5\%)\end{tabular}                        \\ \hline
\multicolumn{1}{c|}{\textbf{Average}}                   & \multicolumn{1}{l|}{8,5}                       & \multicolumn{1}{l|}{3}                        & 7                         & \multicolumn{1}{l|}{4,17}                     & \multicolumn{1}{l|}{0,5}                      & 6,83                     & \multicolumn{1}{l|}{50\%}                      & \multicolumn{1}{l|}{66.67\%}                   & 50\%                      & \multicolumn{1}{l|}{68,66\%}                                                                         & \multicolumn{1}{l|}{81,44\%}                                                                          & 65,14\%                                                                         \\ \cline{2-13} 
\end{tabular}
\end{table*}

\lstset{frame=tb,
  language=Java,
  aboveskip=3mm,
  belowskip=3mm,
  showstringspaces=false,
  columns=flexible,
  basicstyle={\small\ttfamily},
  numbers=left,
  numberstyle=\tiny\color{gray},
  keywordstyle=\color{blue},
  commentstyle=\color{dkgreen},
  stringstyle=\color{mauve},
  breaklines=true,
  breakatwhitespace=true,
  tabsize=3
}


\begin{minipage}{\linewidth}
\begin{lstlisting}[language=Java,label={lst:sample-bard-code},caption=A function generated by Bard for the ``Condominium Mgt.'' assignment that fails our Code Quality validations. Besides using explicit type-testing (i.e.\, uses instanceof)\, Bard also failed to implement the proper business rules\, since the formula for the `Garage' should multiply the area by 3 instead of 4 (see line 3).]
public int calculateCondominiumPayment() {
    int value = baseValue;
    value += area * 4;
    if (this instanceof Apartment) {
        value += ((Apartment) this).getFloorNumber() * 3;
        value += ((Apartment) this).getNrRooms() * 2;
    }
    else if (this instanceof Store) {
        value += ((Store) this).getNrFronts() * 2;
        value += ((Store) this).getNrDoors() * 2;
    }
    else if (this instanceof Garage) {
        value += Math.abs(((Garage) this).getNrFloor()) * 2;
    }
    return value;
}
\end{lstlisting}
\end{minipage}

\subsection{Problem categorization}

This section presents a brief categorization of the problems observed across assignments and models. After each problem, we indicate which models displayed it at least in one assignment.


\textbf{Issues related with syntax}


\begin{itemize}
    \item Failed to include library imports. {\small [GPT-3.5, Bard]}
    \item Created minor compilation errors (e.g. missing ``\}''). {\small [GPT-3.5, Bard]}
    \item Failed to declare a required package. {\small [GPT-3.5]}
    \item Generated incoherent code (e.g. called a function with less arguments than it receives). {\small [Bard]}
\end{itemize}


\textbf{Issues related with OOP concepts}
\begin{itemize}
    \item Used instanceof or getClass() to decide program behaviours. {\small [all]}
    \item Failed to declare a class as abstract. {\small [all]}
    \item Failed to correctly identify and implement business rules. {\small [all]}
    \item Did not respect the `toString()' function's required format. {\small [all]}
    \item Failed to apply inheritance best practices. {\small [all]}
    \item Suggested solutions with code duplication. {\small [all]}
    \item Accessed a super-class’s private fields from its sub-classes. {\small [GPT-3.5, Bard]}
    \item Declared constructors with problems (e.g. added extra arguments). {\small [GPT-3.5]}
    \item Failed to properly apply the downcast operator. {\small [Bard]}
    \item Declared abstract methods in the super-class but failed to implement them in the sub-classes. {\small [GPT-4]}
\end{itemize}






\section{Recommendations for CS Educators}



Students are likely to misuse tools like GPT-3.5 and Bard, which offer decent-to-good OOP code generation for free. Superior models like GPT-4 are currently available at a cost. With the evident progress from GPT-3.5 to GPT-4, observed both in other research \citep{savelka2023thrilled} as well as in our study, we believe that educators should adapt to this emerging landscape.

\subsection{Give more weight to code quality evaluations}


For educators focusing on object-oriented programming, given our observations, as well as previous research \citep{keuning2023systematic}, we recommend putting more weight on evaluating items such as code quality, design patterns and other similar aspects. The course's focus should change from just producing ``functional code'' to producing ``functional and high-quality code''. The assessment work can be scaled using an AAT with the capacity to evaluate functional and quality requirements.


\subsection{Use LLMs in your classes}

Consider embracing LLMs in your classes. One option is the adoption of in-class exercises where students have to 1) interact with LLMs to generate code, and then, 2) evaluate the respective solutions. This process would be supervised by the teacher, who would help the students analyze and critique the LLMs' output. This has the advantage of showing students that these models should not be trusted blindly, and that they should inspect and test the generated code, possibly improving their critical thinking, similarly to other mistake-finding exercises \citep{naumova2023mistake}. This exercise can be enhanced by having students validate their findings through an AAT using teacher-defined tests or by creating their own unit tests.


\subsection{Adopt project-based learning}



If you currently rely solely on using small assignments for assessments, we recommend that you consider including also project-based evaluations. Projects will require more complex code and interactions to be implemented, which will possibly require more complex interactions with the LLMs to obtain a fully working solution that respects the object-oriented best practices. The projects can be incremental, in order to allow students to grow and apply their knowledge incrementally. For example: consider having a first project delivery where the students do not need to use inheritance, followed by another delivery where inheritance is needed, followed by another delivery where some of the requirements change, and so on. Although LLMs' performance over larger assignments has not yet been measured, there are known benefits to project-based learning \citep{mills2003engineering, shin2018effects}.




\section{On the importance of citing your sources}

One of the interesting features of Bard is its ability to display sources for parts of the generated content. In the case of the ``IT Company'' assignment, Bard displayed a single source: a GitHub repository with a Java course.\footnote{https:github.com/camilaabrantes/CursoJava}





We performed a brief analysis of the repository and found some OOP examples that are incorrect. For example, the repository contains a class called `Product' with a `getPrice()' method, which is extended by the sub-class `ImportedProduct' that redefines the price calculation formula to take into account a customs fee. Instead of overriding the `getPrice()', `ImportedProduct' declares a new method `totalPrice()'. This is clearly wrong, in light of OOP best practices. 

Although we are unsure of how much of this repository actually contributed to Bard's reply, we suspect that it is the use of data sources that have not been curated that results in the generation of code that does not follow the OOP best practices.

Note that GPT-3.5 and GPT-4 do not disclose any information regarding their sources. Since the only publicly available information are high-level descriptions of the training datasets (e.g. a filtered version of Common Crawl \footnote{https://commoncrawl.org/the-data}, English Wikipedia, among others) \cite{brown2020language}, it was not possible to perform this analysis for those models.


\section{Limitations}


The 3 experiments were done using different interaction interfaces. This may have some kind of impact on the model’s replies, namely in terms of the number of prompts required to obtain a solution.


These models' output is not deterministic: repeating the experiments for each assignment might have led us to different results.


We did not employ any Prompt Engineering (PE) techniques. The models were initially inputted with the original assignments' text. The prompts used to obtain missing elements or to guide the models' toward better solutions consisted of very literal and direct requests. It is possible that the models would have given better if some PE techniques were applied.

\section{Conclusions and Future Work}

In this work, we present the performance of 3 LLMs when solving several OOP assignments.

From our observations, GPT-4 demonstrated better performance than GPT-3.5, needing fewer interactions and producing code with fewer issues. Additionally, GPT-4 consistently passed more unit tests than GPT-3.5. In comparison, Bard lagged behind both GPT versions in number of issues, passed tests, and the number of interactions needed to guide it toward a reasonable solution.

If the 3 models were students, we would say that GPT-3.5 is a student that has learned the basics of programming but struggles with object-oriented design and programming concepts; GPT-4 is a more experienced student who, with minimal guidance, is able to achieve good solutions; and Bard is a student of a level slightly below GPT-3.5, with less autonomy than the others and more issues with basic programming elements. In the end, all these students struggled somehow and made some mistakes.

But can we blame students for making mistakes when they had a bad teacher? Perhaps the challenges encountered by LLMs in adhering to OOP best practices are not due to inherent reasoning limitations often associated with this technology but rather attributed to suboptimal choices in their information resources. We have verified the poor quality of the source referenced by Bard, and while GPT-3.5 and GPT-4 do not disclose their sources, it is plausible that they too might have relied on inadequate references. The significance of LLMs disclosing their sources, and the more general issue of explaining their reasoning has been a topic of repeated discussion \citep{arrieta2020explainable}. We not only endorse such explanations but also emphasize the necessity of curating high quality sources for training these models.

As for future work, we plan to investigate how these models handle larger OOP assignments (e.g. 10-15 classes) following our recommendation to switch to project-based learning. We also plan on experimenting new ways of presenting OOP exercises to the students, with the goal of creating barriers to naive `copy-and-prompt' approaches. One of the possible approaches is the creation of diagram-based and video-based assignments.


\section{Data Availability}

Since this work is based on real assignments that we use for student assessment in our course, publishing the full assignment and/or interactions dataset would require us to fully recreate all the assignments, ensuring we use unpublished materials when evaluating our students. As such, we have decided to release just one of the assignments: the IT Company Assignment. This was achieved by creating the respective GitHub repository \footnote{https://github.com/drop-project-edu/itCompanyAssignment}, as well as publishing the respective interaction logs, for each LLM, in Zenodo \citep{cipriano_2023_8246165}.

\begin{acks}
This research was funded by the Fundação para a Ciência e a Tecnologia under Grant No.: UIDB/04111/2020 (COPELABS).
\end{acks}

\bibliographystyle{ACM-Reference-Format}
\bibliography{sample-authordraft}



\end{document}